\documentclass[reprint,
					aps,
					pra,
					english,
					footinbib,
					10pt,
					superscriptaddress
]{revtex4-1}

\usepackage[english]{babel}

\usepackage[utf8]{inputenc}

\usepackage[T1]{fontenc}

\usepackage[intlimits]{amsmath}

\usepackage{amssymb}

\usepackage{mathtools}

\usepackage{dsfont}

\usepackage{siunitx}

\usepackage[final,colorlinks=true,linkcolor=blue,citecolor=blue]{hyperref}

\renewcommand*\d[1]{d #1\,}
\newcommand{\ii}{\mathrm{i}}
\DeclarePairedDelimiter{\abs}{\lvert}{\rvert}
\newcommand{\ee}[1]{\operatorname{e}^{#1}}
\newcommand*{\defeq}{\coloneqq}
\newcommand*{\defeqinv}{\eqqcolon}

\DeclareMathOperator{\vspan}{span}
\DeclareMathOperator{\tr}{tr}
\newcommand*{\conj}[1]{{#1}^{*}}

\DeclarePairedDelimiter\ket{\vert}{\rangle}

\DeclarePairedDelimiterX\braket[2]{\langle}{\rangle}{#1\,\delimsize\vert\,\mathopen{}#2}
\DeclarePairedDelimiterX\matrixel[3]{\langle}{\rangle}{#1\,\delimsize\vert\,\mathopen{}#2\,\delimsize\vert\,\mathopen{}#3}
\newcommand{\op}[1]{\hat{#1}}

\begin{document}
\title{Symmetries and entanglement features of inner-mode resolved correlations of interfering nonidentical photons}
\author{Simon Laibacher}
\email{simon.laibacher@uni-ulm.de}
\affiliation{Institut f\"{u}r Quantenphysik and Center for Integrated Quantum Science and Technology (IQ\textsuperscript{ST}), Universität Ulm, D-89069 Ulm, Germany}
\author{Vincenzo Tamma}
\email{vincenzo.tamma@port.ac.uk}
\affiliation{Institut f\"{u}r Quantenphysik and Center for Integrated Quantum Science and Technology (IQ\textsuperscript{ST}), Universität Ulm, D-89069 Ulm, Germany}
\affiliation{School of Mathematics and Physics, University of Portsmouth, Portsmouth PO1 3QL, UK}
\affiliation{Institute of Cosmology \& Gravitation, University of Portsmouth, Portsmouth PO1 3FX, UK}

\begin{abstract}
Multiphoton quantum interference underpins fundamental tests of quantum mechanics and quantum technologies.
Consequently, the detrimental effect of photon distinguishability in multiphoton interference experiments can be catastrophic.
Here, we employ correlation measurements in the photonic inner modes, time or frequency, to restore quantum interference between photons differing in their colors or injection times in arbitrary linear optical networks, without the need for additional filtering or post selection.
Interestingly, we demonstrate how harnessing the multiphoton inner-mode quantum information enables to unravel symmetries of multiphoton networks and states and the generation of an entire class of multipartite entangled states with a fixed interferometer.
These results are therefore of profound interest for future applications of universal inner-mode resolved linear optics across fundamental science and quantum technologies with photons with experimentally different spectral properties.
\end{abstract}

\maketitle

The nonclassical interference of light \cite{Hong1987_Measurementsubpicosecondtime,Alley1986_newtypeEPR,Shih1988_NewTypeEinsteinPodolskyRosenBohm} is one of the main consequences of the quantum nature of the electromagnetic field and lies at the heart of many quantum optics experiments \cite{Hong1987_Measurementsubpicosecondtime,Alley1986_newtypeEPR,Shih1988_NewTypeEinsteinPodolskyRosenBohm,Pan2012_Multiphotonentanglementinterferometry,Tamma2015_MultibosonCorrelationInterferometry,Metcalf2013_Multiphotonquantuminterference,Carolan2015_Universallinearoptics,Flamini2015_Thermallyreconfigurablequantum,HanburyBrown1956_TestNewType,Tamma2014_Multibosoncorrelationinterferometry,Tamma2016_Multipathcorrelationinterference,Cassano2017_Spatialinterferencepairs,Peng2016_ExperimentalcontrolledNOTgate,DAngelo2017_Characterizationtwodistant}.
It plays a central role in a variety of applications ranging from quantum computing \cite{Pan2012_Multiphotonentanglementinterferometry,Ladd2010_Quantumcomputers,Franson2013_BeatingClassicalComputing,Knill2001_schemeefficientquantum,Aaronson2011_computationalcomplexitylinear,Laibacher2015_PhysicsComputationalComplexity} over quantum communication and quantum cryptography \cite{Lo2014_Securequantumkey,Mattle1996_Densecodingexperimental} to quantum metrology and state tomography \cite{Dowling2008_Quantumopticalmetrology,Lemos2014_Quantumimagingundetected,Wasilewski2007_SpectralDensityMatrix,HanburyBrown1956_TestNewType,Tamma2014_Multibosoncorrelationinterferometry,Tamma2016_Multipathcorrelationinterference,Cassano2017_Spatialinterferencepairs,DAngelo2017_Characterizationtwodistant}.
Conventional multiphoton experiments in linear interferometers rely on measurements at the interferometer output which do not resolve the structure of the multiphoton interference in the photonic spectral degrees of freedom, namely frequency and time, effectively ignoring the full quantum information encoded in the photonic spectra.
This ``ignorance'' can lead to the degradation of the observed multiphoton interference with increasing distinguishability for photons with nonidentical input states \cite{Tamma2015_MultibosonCorrelationInterferometry}.
This is indeed the case for single-photon emitters, such as diamond colour centers \cite{Babinec2010_diamondnanowiresinglephoton}, single molecules \cite{Lounis2000_Singlephotonsdemand} and quantum dots \cite{Shields2007_Semiconductorquantumlight,Michler2000_QuantumDotSinglePhoton}.
Here,  photons emitted by different  sources or by the same source at different times are generally different in their spectra.

Fortunately, with the advent of detectors with unprecedented time- or frequency-resolution, linear optical correlation experiments based on inner-mode resolving measurements either in time or frequency have become feasible \cite{Legero2004_QuantumBeatTwo,Avenhaus2009_Fiberassistedsinglephotonspectrograph,Davis2017_Pulsedsinglephotonspectrometer,Polycarpou2012_AdaptiveDetectionArbitrarily,Gerrits2015_Spectralcorrelationmeasurements,Jin2015_SpectrallyresolvedHongOuMandel,Shcheslavskiy2016_Ultrafasttimemeasurements,GrimauPuigibert2017_HeraldedSinglePhotons}.
As a result, full multiphoton interference can be observed at the output of a linear network even in the case of nonidentical input photons \cite{Tamma2015_MultibosonCorrelationInterferometry,Legero2004_QuantumBeatTwo}.
Additionally, the dependence of the correlations of three photons at the output of a linear network has been investigated as a function of the spectral overlap of the input photons \cite{Tan2013_SUQuantumInterferometry,deGuise2014_Coincidencelandscapesthreechannel}.

Furthermore, the generation of maximally entangled W-states was also demonstrated theoretically by postselecting events at equal detection times at the output of a tritter for photons of completely different colors \cite{Tamma2015_MultibosonCorrelationInterferometry}.

It was also shown that the access to the quantum information encoded in the spectra of the interfering photons via correlation measurements in the photonic inner degrees of freedom can unravel the full classical hardness of multiphoton interference in boson sampling schemes \cite{Aaronson2011_computationalcomplexitylinear,Laibacher2015_PhysicsComputationalComplexity}.
This is even the case if no overlap between the input photon frequency and temporal spectra occurs \cite{Laibacher2015_PhysicsComputationalComplexity,Tamma2015_Multibosoncorrelationsampling,Tamma2014_Samplingbosonicqubits}.
Furthermore, it is possible to approach, in principle, deterministic boson sampling realizations with photons of random spectral overlap and/or random input photonic occupation numbers \cite{Laibacher2018_quantumcomputationalsupremacy}.

Despite all these remarkable results, the full quantum advantages of multiphoton interference based on inner-mode resolved linear optics arising in quantum optics experiments even beyond boson sampling are still far from being fully explored.
In particular, working towards novel schemes for the characterization of multiphoton networks and entanglement generation with nonidentical photons, important questions arise:

a) How can given symmetries in the multiphoton input state and in its evolution in a linear optical network be inferred from the measurement of inner-mode correlations at the network output?
b) How do time and frequency resolved measurements tailor the type of entanglement correlations at the output of a linear network depending on the photonic input spectra?

By tackling these fundamental questions, we demonstrate in this paper how the full set of outcomes of inner-mode resolved measurements of multiple nonidentical input photons in a linear optical network allows one to:
a) unravel symmetries of optical networks and of multiphoton quantum states;
b) generate a whole class of entangled multiphoton states even with a fixed configuration of the linear optical network.
Remarkably, these results apply to photons of either different colors or injection times, dramatically increasing the number of possible sources that can be exploited for future experiments (e.g.\ quantum dots).

\begin{figure}[ht]
	\centering
	\includegraphics{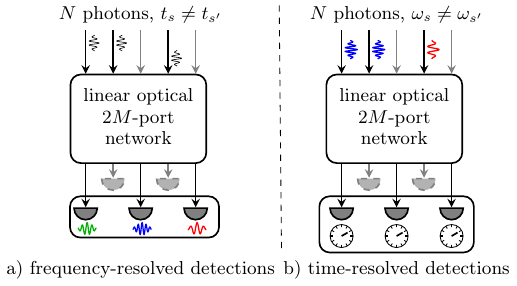}
	\caption{
		a) Setup for frequency-resolved correlation measurements: $N$ single photons, with generally different but overlapping frequency distribution $\xi_s$, are injected at different times $t_s\neq t_{s'}$ ($s,s'\in\mathcal{S}$) into a subset $\mathcal{S}$ of the input ports of a passive linear network and detected using frequency-resolving detectors.
		b) Setup for time-resolved correlation measurements: $N$ single photons with different colors $\omega_s\neq \omega_{s'}$ but overlapping temporal distributions $\mathcal{F}[\xi_s]$ are injected into a passive linear network and detected using time-resolving detectors.
	}
	\label{fig:ScattershotSetup}
\end{figure}

All of the experimental scenarios described in this paper are based on the $N$-photon linear optical networks depicted in Fig.~\ref{fig:ScattershotSetup} with $M\geq N$ ports $s=1,2,...,M$.

Contrary to conventional multiphoton linear optical networks, $N$ input single photons
\begin{equation}
	\ket{1;\xi_s,\omega_s,t_{s}}_{s} \defeq
	\int_{0}^{\infty} \d{\omega} \xi_{s}(\omega-\omega_s) \ee{+\ii \omega t_{s}} \hat{a}_{s}^{\dagger}(\omega) \ket{0}_{s}
	\label{eq:SinglePhotonState}
\end{equation}
with nonidentical normalized spectra $\xi_s (\omega - \omega_s) \ee{\ii \omega t_s}$, differing either in their injection times $\{t_s\} \defeq \{ t_s \in \mathds{R} \mid s \in \mathcal{S} \}$ or in their central frequencies $\{\omega_s\} \defeq \{ \omega_s \in \mathds{R}^+ \mid s \in \mathcal{S} \}$, are injected in a set $\mathcal{S}$ of $N$ input ports, leading to the overall input state
\begin{equation}
	\ket{\mathcal{S}} \defeq \smashoperator[r]{\bigotimes_{s\in \mathcal{S}}}
	\ket{1;\xi_s,\omega_s,t_{s}}_{s}
	\smashoperator[r]{\bigotimes_{s \notin \mathcal{S}}}
	\ket{0}_{s}.
	\label{eq:StateDefinition}
\end{equation}
For simplicity, we assume that the spectra of the input photons satisfy the narrow bandwidth approximation and a polarization-independent interferometric evolution.
Furthermore, given an overall frequency spread $\Delta\omega_{\text{tot}}$ of the input light, we assume that all possible paths through the network are equal on the scale of the coherence length $c/\Delta\omega_{\text{tot}}$.
In this case, the interferometric evolution is also frequency independent and can be described by a single unitary $M\times M$ matrix $\mathcal{U}$ which defines the linear transformation
\begin{equation}
	\label{eq:lineartransformation}
	\hat{a}_d(\omega) = \sum_{s=1}^{M}\mathcal{U}_{ds} \hat{a}_s(\omega)
\end{equation}
between the mode operators $\op{a}_{d}(\omega)$ and $\op{a}_s(\omega)$ at the output and input of the network, respectively \footnote{See section I of the Supplemental Material for details.}.

The $N$ photons are subsequently detected in a subset $\mathcal{D}$ containing $N$ of the output ports $d=1,\dots,M$ and at frequencies $\{\omega_d\}\defeq \{\omega_d \in \mathds{R}^{+}\mid d\in \mathcal{D}\}$ or at times $\{t_d\}\defeq \{t_d \in \mathds{R}\mid d\in \mathcal{D}\}$.
Indeed, these measurements ``erase'' the distinguishability of the input photons in the respective conjugate photonic inner parameters:
If the photons are distinguishable in frequency, multiphoton indistinguishability at the output of the network is ensured by a small enough detector integration in time \cite{Tamma2015_MultibosonCorrelationInterferometry}
\begin{equation*}
	\label{eq:conditions_integration_time}
	\delta t \ll 1/\Delta \omega_{\text{tot}},
\end{equation*}
while a high frequency resolution
\begin{equation*}
	\label{eq:conditions_integration_frequency}
	\delta \omega  \ll \abs{t_{s}-t_{s'}}^{-1} \ \forall s,s'\in \mathcal{S} \quad \text{and} \quad \delta\omega \ll \Delta\omega_{\text{tot}},
\end{equation*}
ensures the indistinguishability of photons injected at different times \cite{Laibacher2018_quantumcomputationalsupremacy}.

In order to emphasize that our results apply to both time and frequency resolved detection schemes due to the conjugacy of time and frequency, we will from here on use the following notation:
The \emph{inner-mode parameter} measured at detector $d$ will be denoted as $\beta_d$ ($\beta_d=t_d$ or $\beta_d=\omega_d$ for time or frequency resolved detection, respectively) while the conjugate inner-mode parameter in which the photons are distinguishable at the input ports $s$ in Eq.~\eqref{eq:StateDefinition} will be labeled as $\alpha_s$ ($\alpha_s=\omega_s$ or $\alpha_s=t_s$ for time or frequency resolved detection, respectively).

The detection probability at the output at frequencies or times $\beta_d = \omega_d,t_d$ for input photons of different central times or frequencies $\alpha_s = t_s,\omega_s$, respectively, can be easily expressed in terms of the bosonic mode operators $\op{a}_d(\omega_d)$ or $\op{a}_d(t_d) \defeq \int d\omega \, \op{a}_d(\omega) \ee{-\ii \omega t_d} / \sqrt{2\pi}$ at the output channels as
\cite{Tamma2015_MultibosonCorrelationInterferometry}
\begin{equation*}
	\label{eq:probability_ideal_measurement_general_state_start}
	\begin{split}
			P^{(\mathcal{D}, \mathcal{S})}_{\{\beta_{d}\},\{\alpha_s\}}
	&\propto \matrixel[\big]{\mathcal{S}}{\smashoperator[r]{\prod_{d \in \mathcal{D}}} \hat{a}^{\dagger}_{d}(\beta_d)
			\smashoperator[r]{\prod_{d \in \mathcal{D}}} \hat{a}_d(\beta_d)}{\mathcal{S}}
	\end{split}
\end{equation*}
Defining $f_s(\beta_d- \beta_s) = \xi_s (\omega_d - \omega_s)$ for frequency resolved detection or $f_s(\beta_d-\beta_s) = \mathcal{F}[\xi_s] (t_d - t_s)$ for time resolved detection (where $\mathcal{F}[\xi_s]$ is the Fourier transform of the frequency distribution $\xi_s$) and using the linear transformation in \eqref{eq:lineartransformation} connecting the mode operators at the input channels $\mathcal{S}$ and the output channels $\mathcal{D}$, this can be rewritten as
\footnote{See section II of the Supplemental Material.}
\begin{equation}
	\label{eq:probability_ideal_measurement_general_state}
	\begin{split}
			P^{(\mathcal{D}, \mathcal{S})}_{\{\beta_{d}\},\{\alpha_s\}}
		&\propto   \abs[\Big]{\sum_{\sigma}\prod_{d\in\mathcal{D}}\mathcal{U}_{d\sigma(d)} f_{\sigma(d)}(\beta_d-\beta_{\sigma(d)})\ee{\ii \beta_{d} \alpha_{\sigma(d)}} }^2.
	\end{split}
\end{equation}
Here, the sum runs over all possible multiphoton paths $\sigma$ (permutations from the symmetric group of order $N$) which bijectively connect the output ports $\mathcal{D}$ with the input ports $\mathcal{S}$.

The probabilities in Eq.~\eqref{eq:probability_ideal_measurement_general_state} are the result of the interference between $N!$ multiphoton probability amplitudes each corresponding to one of the possible multiphoton quantum paths from the sources to the detectors \cite{Tamma2015_MultibosonCorrelationInterferometry,Tamma2015_Multibosoncorrelationsampling}.
These amplitudes are not only determined by the linear network but also depend on the state of the input photons and on the detected frequencies or times.
This is a manifestation of the drastically enlarged Hilbert space accessible by inner-mode resolved detections which can be employed as a quantum resource to unravel symmetry structures in multiphoton interference patterns as well as to tailor non-local $N$-photon correlations.

\paragraph{Symmetries of inner-mode resolved correlations.}
We show how symmetries in the interference pattern of the correlations in the photonic inner modes described by Eq.~\eqref{eq:probability_ideal_measurement_general_state} provide a powerful tool to reveal information about the $N$-photon states, their interferometric evolution, or both simultaneously.
\begin{figure}[ht]
	\centering
	\includegraphics{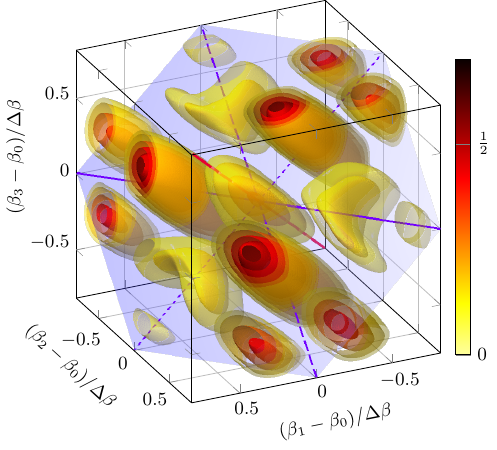}
	\caption{The inner-mode resolved interference pattern defined by Eq.~\eqref{eq:probability_ideal_measurement_general_state} at the output of a symmetric tritter for three photons with input parameters $\{\alpha_{s}\}=\{0,7.4/\Delta\beta,11.3/\Delta\beta\}$ ($\alpha_s=t_s,\omega_s$) and equal Gaussian distributions $f(\beta_d-\beta_0)$ of bandwidth $\Delta\beta$, centered at $\beta_0$ ($\beta_d=\omega_d,t_d$).
		The behaviour of the multiphoton interference pattern under permutations $\tau$ of the detected inner-mode values $\beta_d$ reveals two distinct classes of symmetries: a) a threefold rotational symmetry around the axis $(1,1,1)$ %
		[{\protect\includegraphics{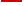}}]
		if $\tau$ is $(1)(2)(3)$ [rotation by \SI{0}{\degree}], $(123)$ [\SI{120}{\degree}], or $(132)$ [\SI{240}{\degree}], arising uniquely from the symmetric tritter, described by Eq.~\eqref{eq:symmetry_permutation};
		b) three twofold rotational symmetry axes which correspond to $\tau$ being $(12)(3)$ %
		[{\protect\includegraphics{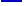}}],
		$(23)(1)$ %
		[{\protect\includegraphics{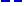}}],
		or $(13)(2)$
		[{\protect\includegraphics{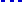}}],
		emerging from a combination of the network's symmetry and a symmetry of the input frequency distributions around a common central frequency, described by Eq.~\eqref{eq:symmetry_permutation_and_parity}.
		Lastly, a symmetry plane (light blue) orthogonal to the axis $(1,1,1)$ exists due to the highly symmetric input state of the photons (equal Gaussian inner-mode distributions), independently of the unitary transformation incurred by the linear network.
}
	\label{fig:correlation_landscape}
\end{figure}
Indeed, each $N$-photon interference amplitude in these correlations is given by the product of an interferometric amplitude $\mathcal{A}_{\sigma}\defeq \prod_{d\in\mathcal{D}}\mathcal{U}_{d\sigma(d)}$ and a spectral amplitude $\mathcal{B}_{\sigma}(\{\beta_d\},\{\alpha_s\})\defeq \prod_{d\in\mathcal{D}} f_{\sigma(d)}(\beta_d - \beta_{\sigma(d)})\ee{\ii\beta_d \alpha_{\sigma(d)}}$.
To investigate the symmetry properties of the $N$-photon detection probability in Eq.~\eqref{eq:probability_ideal_measurement_general_state}, we consider its behaviour under an arbitrary linear transformation $\mathcal{T}: \{\beta_d\} \to \vspan(\{\beta_d\})^{\otimes N}$ of its inner-mode arguments $\beta_d$ leading to
\begin{equation}
	\label{eq:general_transformation_of_innermodes}
	P^{(\mathcal{D},\mathcal{S})}_{\mathcal{T}(\{\beta_d\}),\{\alpha_s\}}
	= \abs[\Big]{\sum_{\sigma} \mathcal{A}_{\sigma} \mathcal{B}_{\sigma}\bigl(\mathcal{T}(\{\beta_d\}),\{\alpha_s\}\bigr)}^2.
\end{equation}

As a first example, we address the case where the multiphoton spectral amplitude $\mathcal{B}_{\sigma}$ defined by the spectra of the input photons is symmetric under a given transformation $\mathcal{T}$ apart from a permutation-independent factor, i.e. $\mathcal{B}_{\sigma}\bigl(\mathcal{T}(\{\beta_d\}),\{\alpha_s\}\bigr) = \text{const.} \times  \mathcal{B}_{\sigma}(\{\beta_{d}\},\{\alpha_s\})$.
In this case also the correlations in Eq.~\eqref{eq:general_transformation_of_innermodes} manifest the same symmetry.
Therefore, symmetric properties of the input temporal and frequency spectra can be revealed from the measured symmetries of the multiphoton interference pattern independently of the optical network.
For example, as we will show later, a mirror symmetry of the measured inner-mode resolved correlations $P$ can arise from the highly symmetric input state of photons with the same Gaussian distribution $f(\beta_d - \beta_0)$ centered at the frequency or time $\beta_0 = \omega_0, t_0$, respectively.

As a second example, we consider photonic input spectral distributions $f_s(\beta_d-\beta_0)$ in general different but symmetric around a common value $\beta_0$.
In this case, the multiphoton amplitudes satisfy the property
\begin{equation}
	\label{eq:conjugate_symmetry_spectral_amplitude}
	\conj{\mathcal{B}}_{\sigma}(\{\beta_d\},\{\alpha_s\}) =  \mathcal{B}_{\sigma}(\{2\beta_0-\beta_d\},\{\alpha_s\}),
\end{equation}
apart from a phase term independent of the permutation $\sigma$.
Consequently, the measured multiphoton interference pattern in Eq.~\eqref{eq:general_transformation_of_innermodes} is invariant under the parity transformation $\mathcal{T}: \beta_d -\beta_0 \mapsto -(\beta_d-\beta_0)$ if the unitary interferometer transformation is characterized by multiphoton amplitudes $\mathcal{A}_{\sigma}$ with the same phase $\ee{\ii\varphi}$ \footnote{See section IIIa of the Supplemental Material}.

As a final example, we consider the case of a permutation $\tau$ of the inner-mode arguments in Eq.~\eqref{eq:general_transformation_of_innermodes} corresponding to the transformation $\mathcal{T}_{\tau}: \beta_d \mapsto \beta_{\tau(d)}$.
By reordering the product defining the spectral amplitude, it is straightforward to show that $\mathcal{B}_{\sigma}(\{\beta_{\tau(d)}\},\{\alpha_s\}) = \mathcal{B}_{\sigma\circ\tau^{-1}}(\{\beta_d\},\{\alpha_s\})$, where $\circ$ denotes the concatenation of permutations.
Consequently, a permutation of the values $\beta_d$ can be mapped to a permutation of the interferometric amplitudes $\mathcal{A}_{\sigma}$ as \footnote{See section IIIb of the Supplemental Material.}
\begin{equation}
	\label{eq:permutation_of_detected_frequencies}
	P^{(\mathcal{D},\mathcal{S})}_{\{\beta_{\tau(d)}\},\{\alpha_s\}} = \abs[\Big]{\sum_{\sigma}\mathcal{A}_{\sigma\circ\tau}\mathcal{B}_{\sigma}(\{\beta_d\},\{\alpha_s\})}^2.
\end{equation}
Interestingly, this implies that some symmetries in the measured inner-mode resolved correlations can be intrinsically connected to symmetries in the interferometric amplitudes.
Namely, if $\mathcal{A}_{\sigma\circ\tau}=\mathcal{A}_{\sigma} \ee{\ii\varphi} \forall\sigma$ with a phase $\varphi$ for a given permutation $\tau$ the interference pattern is symmetric under the corresponding permutation of the detected parameters $\{\beta_d\}$,
\begin{equation}
	\label{eq:symmetry_permutation}
	P^{(\mathcal{D},\mathcal{S})}_{\{\beta_{\tau(d)}\},\{\alpha_s\}} = P^{(\mathcal{D},\mathcal{S})}_{\{\beta_{d}\},\{\alpha_s\}}.
\end{equation}
Furthermore, in the case where $\mathcal{A}_{\sigma\circ\tau}=\conj{\mathcal{A}}_{\sigma} \ee{\ii\varphi} \forall\sigma$, we find -- using Eqs.~\eqref{eq:conjugate_symmetry_spectral_amplitude} and \eqref{eq:permutation_of_detected_frequencies} -- that the correlations are symmetric under a combination of the permutation $\tau$ and the parity operation $\beta_d \rightarrow 2\beta_0-\beta_d$ if the spectral distributions are all symmetric around the same central value $\beta_0$.
\begin{equation}
	\label{eq:symmetry_permutation_and_parity}
	P^{(\mathcal{D},\mathcal{S})}_{\{2\beta_0-\beta_{\tau(d)}\},\{\alpha_s\}} = P^{(\mathcal{D},\mathcal{S})}_{\{\beta_{d}\},\{\alpha_s\}}
\end{equation}

In order to provide a practical example of some of the symmetric properties described so far, we consider the case of three photons with identical spectral distributions $f(\beta_d-\beta_0)$ but different injection times or central frequencies $\alpha_s=t_s,\omega_s$ measured at the output of a symmetric tritter at frequencies $\beta_d=\omega_d$ or times $\beta_d=t_d$, respectively.
The corresponding correlations depicted in Fig.~\ref{fig:correlation_landscape} depicted as a function of the detected inner-mode parameters $\{\beta_d\}$ exhibit not only three-photon quantum beating \cite{Tamma2015_MultibosonCorrelationInterferometry} but also several evident symmetries in the measured photonic inner parameters.
We will now describe how these symmetries originate from either the interferometer transformation $\mathcal{U}$, or the input state, or a combination of both.

We first find from the symmetric tritter single-photon amplitudes $\mathcal{U}_{ds}\defeq \exp(\ii \frac{2\pi}{3}ds)/\sqrt{3}$ ($d,s=1,2,3$) that $\mathcal{A}_{\sigma\circ\tau}=\mathcal{A}_{\sigma}$ if the permutation $\tau$ is either $(1)(2)(3)$, $(123)$, or $(132)$ [using cycle notation of permutations] and consequently that the probability remains unchanged under these permutations of the inner-mode variables $\beta_d$, as is clear from the result for arbitrary values of $N$ in Eq.~\eqref{eq:symmetry_permutation}.
As depicted in Fig.~\ref{fig:correlation_landscape}, this results in a threefold rotational symmetry axis (red line) emerging solely from the symmetry of the linear optical network.
We understand this by noting that the permutation $(123)$ acting onto the vector $(\beta_1,\beta_2,\beta_3)$ corresponds to a permutation matrix
\begin{equation*}
	R_{(123)}=
	\begin{pmatrix}
		0 & 1 & 0 \\
		0 & 0 & 1 \\
		1 & 0 & 0
	\end{pmatrix},
\end{equation*}
which at the same time represents a rotation of $\SI{240}{\degree}$ around the axis $\beta_1=\beta_2=\beta_3$.
Consequently, $\tau=(123)$ yields a rotation of the correlation pattern itself $\SI{-240}{\degree}$ or equivalently $\SI{120}{\degree}$.
Equivalently, the permutation $(132)$ corresponds to a rotation by \SI{240}{\degree} around the same axis.

A second class of symmetries correspond to the permutations $(12)(3)$, $(23)(1)$, or $(13)(2)$ for which $\mathcal{A}_{\sigma\circ\tau}=\conj{\mathcal{A}}_{\sigma}$.
According to Eq.~\eqref{eq:symmetry_permutation_and_parity}, these correlations are symmetric under a combination of these permutations with a parity operation, given a symmetry of the inner-mode distributions of the input photons around a common value of the central inner-mode parameter $\beta_0=\omega_0,t_0$.
These symmetries show up as three distinct twofold rotational symmetry axes in Fig.~\ref{fig:correlation_landscape} (blue lines).
For example, the permutation $(12)(3)$ together with the parity operation is represented by the negative permutation matrix
\begin{equation*}
	-R_{(12)(3)}=
	\begin{pmatrix}
		0 & -1 & 0 \\
		-1 & 0 & 0 \\
		0 & 0 & -1
	\end{pmatrix}
\end{equation*}
which is equivalent to a rotation by $\SI{180}{\degree}$ around the axis $\nu_1+\nu_2=0,\nu_3=0$ (solid blue line in Fig.~\ref{fig:correlation_landscape}).
The two remaining permutations $(23)(1)$ and $(13)(2)$ are analogously connected to twofold rotational symmetries around the axes defined by $\nu_2+\nu_3=0,\nu_1=0$ (dashed blue line) or by $\nu_1+\nu_3=0,\nu_2=0$ (dotted blue line), respectively.

Finally, a mirror symmetry with respect to the blue plane orthogonal to the axis $(1,1,1)$ in Fig.~\ref{fig:correlation_landscape} can be attributed solely to the choice of a highly symmetric input state (three photons with identical Gaussian frequency distributions) \footnote{See section IIIc of the Supplemental Material.}.
In combination with this mirror symmetry, the rotational symmetries corresponding to the permutations $(12)(3)$, $(23)(1)$, and $(13)(2)$ are equivalent to three distinct mirror planes, each spanned by the red axis and one of the blue axes.

We also notice that no parity invariance arises in the interference pattern in Fig.~\ref{fig:correlation_landscape} since the multiphoton amplitudes $\mathcal{A}_{\sigma}$ do not have the same complex phase and therefore do not satisfy the condition pointed out before for parity invariance.

We emphasize that the symmetries in the optical network described here are encoded
in the beating pattern of the correlations in the detected inner-mode values $\beta_d=\omega_d,t_d$
\emph{only} for input photons with different central inner-mode values $\alpha_s=t_s,\omega_s$, respectively.
However, the exact values of the differences $\alpha_s - \alpha_{s'}$ ($s,s'\in\mathcal{S}$) solely determine the periodicity of the observed beatings.
If no inner-mode resolved measurements are employed interference beatings cannot be observed and the corresponding symmetries cannot be retrieved \cite{Tamma2015_MultibosonCorrelationInterferometry,Tamma2015_Bosonsamplingnonidentical}.

\paragraph{Multiphoton entanglement features.}

Remarkably, inner-mode correlation measurements also allow us to encode a whole family of entangled $N$-qubit states in the outcomes of the measurements.
In particular, we will show how entanglement in the polarization modes can be tailored depending on the inner-mode input parameters $\alpha_s$ and the detected conjugate parameters $\beta_d$.
For this purpose, we generalize the photonic input state in Eq.~\eqref{eq:SinglePhotonState} to describe photonic qubits with horizontal (H) or vertical (V) polarization as
\begin{equation}
	\label{eq:singlephotonstate_polarization}
	\ket{1;\xi_s,\omega_s,t_{s},\lambda_s}_{s} \defeq
	\int_{0}^{\infty} \d{\omega} \xi_{s}(\omega-\omega_s) \ee{+\ii \omega t_{s}} \hat{a}_{s,\lambda_s}^{\dagger}(\omega) \ket{0}_{s},
\end{equation}
with $\lambda_s=\text{H},\text{V}$.
Then, inner-mode correlation measurements at the output of a generalised symmetric beam splitter -- described by the unitary $\mathcal{U}_{ds}=\exp(\ii \frac{2\pi}{N}d s)/\sqrt{N}$ and assumed to be polarization independent -- can lead to entanglement correlations spanning the full class of $N$-qubit W-states \cite{Dur2000_Threequbitscan}.
This can be achieved by choosing $N-1$ $H$- and one $V$-polarized photons ($\lambda_1=\dots=\lambda_{N-1}=\text{H}, \lambda_N=\text{V}$) with identical frequency distribution $\xi(\omega)$ as the input state $\bigotimes_{s=1}^{N}\ket{1; \xi, \omega_s, t_s, \lambda_s}_s$.
By propagating this input state to the output of the linear network using Eq.~\eqref{eq:lineartransformation}, we find that a frequency-resolved, $N$-fold coincidence measurement at the output of the network is only sensitive to the contribution \footnote{See section IVa of the Supplemental Material}
\begin{equation*}
	\label{eq:output_state_polarization_no_bunching}%
	\sum_{\sigma\in\Sigma_N}\prod_{d} \mathcal{U}_{d\sigma(d)} \int \d{\beta_d} f_{\sigma(d)}(\beta_{d})\ee{\ii\beta_{d} \alpha_{\sigma(d)}} \ket{\{\lambda_{\sigma(d)}\},\{\beta_d\}}
\end{equation*}
from the overall output state.
For given detected inner-mode values $\{\beta_d\}$, the sum in this expression defines a state from the class of $N$-qubit W-states.

To illustrate this further, we will now consider the case of $N=3$ as an example.
Here, for a given set of measured values $\{\beta_d\}$, the polarization state of the photons corresponds to a three-photon W-state
\begin{equation}
	\label{eq:wstate_definition}
	\ket{W_{\{\beta_d\},\{\alpha_s\}}} \defeq a\ket{\text{HHV}} + b\ket{\text{HVH}} + c\ket{\text{VHH}},
\end{equation}
with amplitudes $a=a_{\{\beta_d\},\{\alpha_s\}}$, $b=b_{\{\beta_d\},\{\alpha_s\}}$, $c=c_{\{\beta_d\},\{\alpha_s\}}$ parametrized by the detected inner-mode values $\{\beta_d\}=\{\omega_d\},\{t_d\}$ and the input values $\{\alpha_s\}=\{t_s\},\{\omega_s\}$, respectively.
As we demonstrate in the Supplemental Material \footnote{See section IVb of the Supplemental Material}, the absolute values are given by
\begin{equation}
	\label{eq:wstateamplitudes}
	\begin{split}
		\abs{a} &= \abs[\Big]{\cos\Bigl( \frac{1}{2}(\beta_2-\beta_3)(\alpha_1-\alpha_2)- \frac{2\pi}{3} \Bigr)},
		\\
		\abs{b} &= \abs[\Big]{\cos \Bigl( \frac{1}{2}(\beta_1-\beta_3)(\alpha_{1}-\alpha_{2}) + \frac{2\pi}{3}\Bigr)},
		\\
		\abs{c} &= \abs[\Big]{\cos \Bigl( \frac{1}{2}\bigl((\beta_1-\beta_3)-(\beta_2-\beta_3)\bigr)(\alpha_{1}-\alpha_{2}) - \frac{2\pi}{3}\Bigr)},
	\end{split}
\end{equation}
apart from a common normalization constant.
We notice that the moduli of these coefficients only arise from the interference between the H-polarized photons and that the degree of entanglement is therefore independent of the inner-mode parameter $\alpha_3$ of the V-polarized photon.
Indeed, if the two $H$-polarized photons are offset in their initial inner-mode parameters with respect to each other ($\alpha_{1}-\alpha_{2} \neq 0$), their interference manifests in a beating behavior of the coefficients $a$, $b$, $c$, independently of the photonic frequency distribution $\xi(\omega)$.
\begin{figure}
	\begin{center}
		\includegraphics{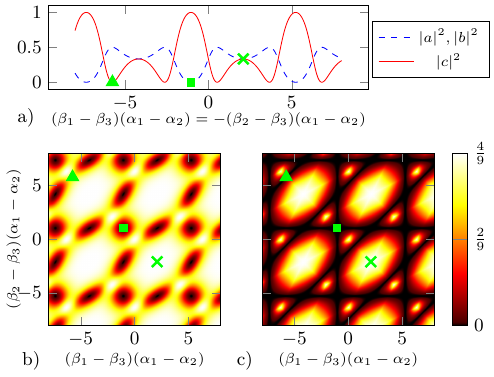}
		\caption{
			a) Beating in the probabilities $\abs{a}^2$, $\abs{b}^2$, $\abs{c}^2$ defining the detected W-state in Eq.~\eqref{eq:wstate_definition} generated by inner-mode resolved correlation measurements in the values $\beta_d=\omega_d,t_d$ of two $H$- and one $V$-polarized input photons of different parameters $\alpha_s=t_s,\omega_s$ at the output of a symmetric tritter. The inner-mode values $\beta$ and $\alpha$ correspond to the frequency and time or vice versa.
			b) Average two-photon concurrence $E^{(\text{av})}_{\{\beta_d\},\{\alpha_s\}}$ in Eq.~\eqref{eq:averageconcurrence}.
			c) Minimum two-photon concurrence $E^{(\text{min})}_{\{\beta_d\},\{\alpha_s\}}$ in Eq.~\eqref{eq:minimalconcurrence}.
			As an example, panel a) is plotted as a function of $(\beta_1-\beta_3)(\alpha_1-\alpha_2) = -(\beta_2-\beta_3)(\alpha_1-\alpha_2)$, corresponding to the antidiagonal of panels b) and c).
			Any degree of entanglement of the W-state class emerges from the beating of the amplitudes $a$, $b$, $c$ caused by a difference in the input inner-mode parameters of the H-polarized photons ($\alpha_1\neq\alpha_2$) when measurements in the conjugate parameters $\beta_d$ are performed.
			As an example, tripartite entanglement of the W-state type, corresponding to the maximal possible value $E_{\text{av}}=E_{\text{min}}=4/9$ \cite{Dur2000_Threequbitscan}, is uniquely achieved for $\abs{a}=\abs{b}=\abs{c} = 1/\sqrt{3}$ (see cross mark as an example).
			Further, the detected state is fully separable when $E_{\text{av}} = E_{\text{min}}=0$ since two of the amplitudes $a$, $b$, and $c$ vanish (e.g. square mark). Last, a biseparable state is found when only one of the amplitudes vanishes (e.g. triangle mark).
	Such a state shows a high average entanglement $E_{\text{av}}=1/3$ since two photons are maximally entangled, but a vanishing minimal entanglement $E_{\text{min}}=0$ due to the biseparability.}
		\label{fig:entanglement}
	\end{center}
\end{figure}
This beating behavior is depicted in Fig.~\ref{fig:entanglement}a) in the particular case $(\beta_1-\beta_3)(\alpha_1- \alpha_2) = - (\beta_2-\beta_3)(\alpha_1-\alpha_2)$ for which $\abs{a}=\abs{b}$.
Therefore, the inner-mode measurements span the full family of three-photon W-states in Eq.~\eqref{eq:wstate_definition} through the beating behavior of the corresponding weights in Eq.~\eqref{eq:wstateamplitudes}
as a function of the detected inner-mode values $\{\beta_d\}$ and the input inner-mode values $\alpha_1-\alpha_2$.
In this respect, the input values $\alpha_s$ can also be spanned experimentally via inner-mode multiplexing with $N$ SPDC sources, independently of the output values $\{\beta_d\}$ \cite{Laibacher2018_quantumcomputationalsupremacy}.

It is straightforward to describe how these inner-mode correlation measurements tailor the W-state entanglement by measuring the corresponding squared concurrences for each of the three possible reduced density matrices obtained by tracing out either photon 1, 2, or 3 \cite{Coffman2000_Distributedentanglement}.
In particular, the average squared concurrence \cite{Dur2000_Threequbitscan}
\begin{equation}
	\label{eq:averageconcurrence}
	E^{(\text{av})}_{\{\beta_d\},\{\alpha_s\}} = \frac{4}{3} \left( \abs{ab}^2 + \abs{bc}^2 + \abs{ac}^2 \right) \leq \frac{4}{9}
\end{equation}
and the minimum squared concurrence
\begin{equation}
	\label{eq:minimalconcurrence}
	E^{(\text{min})}_{\{\beta_d\},\{\alpha_s\}} = 4 \min \left( \abs{ab}^2, \abs{bc}^2, \abs{ac}^2 \right) \leq \frac{4}{9},
\end{equation}
depicted in Fig.~\ref{fig:entanglement}b) and \ref{fig:entanglement}c), respectively, define the degree of entanglement for each W-state as a function of the relative detected inner-mode parameters $\{\beta_d\}$ and of the input inner-mode parameters $\alpha_1-\alpha_2$ according to Eq.~\eqref{eq:wstateamplitudes}.
Evidently, the states encoded into the outcomes of the inner-mode measurements exhibit arbitrary degrees of tripartite entanglement of the W-state type from complete separability up to maximal entanglement (Fig.~\ref{fig:entanglement}) \footnote{See section IVc of the Supplemental Material.}.

\paragraph{Discussion}
We have shown how frequency and time resolved multiphoton interference between nonidentical photons is a promising tool to unravel the symmetries characterizing their quantum states and their evolution in linear optical networks.
Exploiting inner-mode correlation measurements, the differences in the photonic inner-mode parameters (i.e. color or injection times), instead of being a challenge to overcome, become also a powerful resource to generate the entire family of $N$-photon W-states with a single network.
Indeed, this is possible by recording the inner-mode quantum information encoded in the beating behavior of interfering nonidentical photons.

In conclusion, these results have the potential to inspire novel platforms for the analysis of multiphoton linear networks and for multiphoton entanglement generation by employing the full quantum capabilities of inner-mode multiphoton interference in universal linear optics with arbitrary sources of nonidentical photons.
In particular, the experimental verification of the emergence of some of the multiphoton symmetries of optical networks predicted here has just recently been reported in Ref.~\cite{Wang2018_Timeresolvedbosonsampling}.

\begin{acknowledgments}
The authors are grateful to C. Dewdney, S. Kolthammer, A. Laing, F. Sciarrino, and P. Walther for discussions related to this work.

Research was partially sponsored by the Army Research Laboratory and was accomplished under Cooperative Agreement Number W911NF-17-2-0179.
The views and conclusions contained in this document are those of the authors and should not be interpreted as representing the social policies, either expressed or implied, of the Army Research Laboratory or the U.S. Government.
The U.S. Government is authorized to reproduce and distribute reprints for Government purposes notwithstanding any copyright notation herein.
V. T. is also thankful to W. P. Schleich for the time passed until the summer of 2016 at the Institute of Quantum Physics in Ulm where some of the ideas behind this work started to flourish.
S. L. acknowledges support by a grant from the Ministry of Science, Research and the Arts of Baden-W\"urttemberg (Az: 33-7533-30-10/19/2).

Both authors contributed equally to the obtained results.
The project was conceived and managed by V.T.
\end{acknowledgments}

\clearpage

\widetext
\begin{center}
\textbf{\large Symmetries and entanglement features of inner-mode resolved correlations of interfering nonidentical photons: Supplemental Material}
\end{center}
\setcounter{equation}{0}
\setcounter{figure}{0}
\setcounter{table}{0}
\setcounter{page}{1}
\makeatletter
\renewcommand{\theequation}{S.\arabic{equation}}
\renewcommand{\thefigure}{S.\arabic{figure}}
\renewcommand{\bibnumfmt}[1]{[S.#1]}
\renewcommand{\citenumfont}[1]{S.#1}
\pagenumbering{Roman}

\section{Frequency independence of linear networks}

In order to implement a phase shift $0\leq \phi \leq 2\pi$ at the central frequency $\omega_0$ of the light, a change of the optical path length by $\Delta_{\phi} = \phi\, c /\omega_0 \leq 2\pi c/\omega_0$ is sufficient.
This length is on the order of the optical wavelength and can therefore be neglected with respect to all optical path lengths $\Delta x_{ds}^{(i)}$, $\Delta_{\phi}  \ll \Delta x_{ds}^{(i)}$.
Furthermore, if the light has a narrow overall bandwidth $\Delta\omega_{\text{tot}} \ll \omega_0$, $\Delta_{\phi}$ can also be neglected with respect to the correlation length $2\pi c/\Delta\omega_{\text{tot}}$ of the light pulses since
\begin{equation}
	\label{eq:app_bsp:frequencyindependenceinterferometers:smallphasedeviation}
	\frac{\Delta_{\phi}}{2\pi c/\Delta\omega_{\text{tot}}} = \frac{\Delta\omega_{\text{tot}}}{\omega_0} \frac{\phi}{2\pi} \leq \frac{\Delta\omega_{\text{tot}}}{\omega_0} \ll 1.
\end{equation}
Analogously, the frequency dependence of the beam splitters can also be assumed for sufficiently narrow bandwidth.

Let us now turn to the unitary transformation of the linear network.
In typical designs, the beam splitters are located on a regular grid ensuring that the path lengths $\Delta x_{ds}^{(i)}$ of all possible paths through the interferometer are approximately equal to a given length $\Delta x$ on the scale of the correlation length of the light, $\Delta x_{ds}^{(i)} = \Delta x + \delta x_{ds}^{(i)}$ with $\abs{\delta x_{ds}^{(i)}} \ll 2\pi c/\Delta\omega_{\text{tot}} \ \forall i \forall s \forall d$ ($i$ labels all possible paths connecting a fixed pair $s,d$ of input and output channel).
Consequently, with $\omega= \omega_0 + \Omega$ ($\abs{\Omega}\leq \Delta\omega_{\text{tot}}$)
\begin{equation}
	\ee{\ii\omega \Delta x_{ds}^{(i)}/c}
	= \ee{\ii\omega \Delta x/c + \ii\omega_0 \delta x_{ds}^{(i)}/c + \ii\Omega \delta x_{ds}^{(i)}/c}
	\approx \ee{\ii\omega \Delta x/c + \ii\omega_0 \delta x_{ds}^{(i)}/c}
\end{equation}
since
\begin{equation}
	\abs{\Omega} \abs{\delta x_{ds}^{(i)}} \leq \Delta\omega_{\text{tot}} \abs{\delta x_{ds}^{(i)}} \ll c.
\end{equation}
It immediately follows that the total probability amplitude connecting input channel $s$ with output channel $d$ can be written as
\begin{equation}
	\label{eq:app_bsp:frequencyindependenceinterferometers:totalamplitudelayout}
	\mathcal{U}_{d s}(\omega) = \sum_{i} \mathcal{U}_{d s}^{(i)} \ee{\ii \omega \Delta x_{d s}^{(i)}/c} = \ee{\ii \omega \Delta x/c} \sum_{i} \mathcal{U}_{d s}^{(i)} \ee{\ii \omega_0 \delta x_{ds}^{(i)}}= \mathcal{U}_{d s} \ee{\ii\omega \Delta x/c},
\end{equation}
where $\mathcal{U}_{ds}=\sum_{i}\mathcal{U}_{ds}^{(i)}\ee{\ii \omega_0 \delta x_{ds}^{(i)}}$.
Without losing generality, we can set $\Delta x=0$ since non-zero values only correspond to an offset of the detection times $t_d$.

\section{Inner-mode correlations of \texorpdfstring{$N$}{N} photons}

As discussed in the main paper, the probability for an $N$-fold, inner-mode resolved coincidence detection of the photons at the output ports $\mathcal{D}$ and at times or frequencies $\{\beta_d\}=\{t_d\},\{\omega_d\}$ is given by
\begin{equation}
	\label{eq:supp:probability_frequency}
	P^{(\mathcal{D}, \mathcal{S})}_{\{\beta_{d}\},\{\alpha_s\}} =
	\int_{I(\{\beta_d\})}\prod_{d\in\mathcal{D}}\d{\beta_d}\matrixel[\big]{\mathcal{S}}{\smashoperator[r]{\prod_{d \in \mathcal{D}}} \hat{a}^{\dagger}_{d}(\beta_d)
	\smashoperator[r]{\prod_{d \in \mathcal{D}}} \hat{a}_d(\beta_d)}{\mathcal{S}},
\end{equation}
where $\ket{\mathcal{S}}$ is the $N$-photon input state defined in Eq.~\eqref{eq:StateDefinition} of the main paper and $I(\{\beta_d\})\defeq \bigotimes_{d\in\mathcal{D}}[\beta_d -\delta\beta/2, \beta_d +\delta\beta/2]$ are the detector integration intervals.

The correlation function in this expression can be evaluated by using the linear transformation connecting the mode operators of the input and output channels, Eq.~\eqref{eq:lineartransformation} of the main paper.
For a given set $\mathcal{S}$ of input channels, the latter equation effectively becomes
\begin{equation}
	\label{eq:supp:lineartransformation}
	\hat{a}_d(\beta_d) = \sum_{s\in \mathcal{S}}\mathcal{U}_{ds} \hat{a}_s(\beta_d)
\end{equation}
since the unoccupied input ports $s\not\in \mathcal{S}$ do not contribute to the correlations in Eq.~\eqref{eq:supp:probability_frequency}.

Using the notation $\|\ket{\psi}\|^2  \defeq \braket{\psi}{\psi}$ and Eq.~\eqref{eq:supp:lineartransformation}, we can rewrite the correlation function in Eq.~\eqref{eq:supp:probability_frequency} as
\begin{equation}
			\matrixel[\big]{\mathcal{S}}{\smashoperator[r]{\prod_{d \in \mathcal{D}}} \hat{a}^{\dagger}_{d}(\beta_d)
			\smashoperator[r]{\prod_{d \in \mathcal{D}}} \hat{a}_d(\beta_d)}{\mathcal{S}}
		     = \left\| \prod_{d \in \mathcal{D}} \hat{a}_d(\beta_d) \ket{\mathcal{S}} \right\|^2
		     = \left\| \prod_{d\in \mathcal{D}} \sum_{s_d \in \mathcal{S}} \mathcal{U}_{ds_d} \hat{a}_{s_d}(\beta_d) \ket{\mathcal{S}} \right\|^2
		     = \Biggl\| \sum_{\{s_d\}\in \mathcal{S}^{N}} \prod_{d \in \mathcal{D}} \mathcal{U}_{d s_d} \hat{a}_{s_d}(\beta_d) \ket{\mathcal{S}} \Biggr\|^2.
\end{equation}
This expression can be further simplified by noting that due to the structure of the state $\ket{\mathcal{S}}$, Eq.~\eqref{eq:StateDefinition} in the main paper, only those terms contribute, in which each of the $N$ annihilation operators $\hat{a}_{s}(\beta)$, $s\in\mathcal{S}$, appears exactly once.
Denoting the set of all permutations of $N$ elements, the symmetric group of order $N$, as $\Sigma_N$ and recalling Eq.~\eqref{eq:StateDefinition} in the main paper, the correlation function becomes
\begin{equation}
			\matrixel[\big]{\mathcal{S}}{\smashoperator[r]{\prod_{d \in \mathcal{D}}} \hat{a}^{\dagger}_{d}(\beta_d)
			\smashoperator[r]{\prod_{d \in \mathcal{D}}} \hat{a}_d(\beta_d)}{\mathcal{S}}
		     = \left\| \sum_{\sigma \in \Sigma_N} \prod_{d \in \mathcal{D}} \mathcal{U}_{d \sigma(d)} \hat{a}_{\sigma(d)}(\beta_d) \ket{\mathcal{S}} \right\|^2
		     = \left\| \sum_{\sigma \in \Sigma_N} \prod_{d \in \mathcal{D}} \mathcal{U}_{d \sigma(d)} \hat{a}_{\sigma(d)}(\beta_d) \ket{1;\xi_{\sigma(d)},\omega_{\sigma(d)},t_{\sigma(d)}]}_{\sigma(d)} \right\|^2.
\end{equation}
With the help of the definition of the single-photon states $\ket{1;\xi_s,\omega_s,t_s}$ in Eq.~\eqref{eq:SinglePhotonState} of the main paper, this can finally be simplified to
\begin{equation}
	\begin{split}
			\matrixel[\big]{\mathcal{S}}{\smashoperator[r]{\prod_{d \in \mathcal{D}}} \hat{a}^{\dagger}_{d}(\beta_d)
			\smashoperator[r]{\prod_{d \in \mathcal{D}}} \hat{a}_d(\beta_d)}{\mathcal{S}}
			&= \left\| \sum_{\sigma \in \Sigma_N} \prod_{d \in \mathcal{D}} \mathcal{U}_{d \sigma(d)} f(\beta_d-\beta_{\sigma(d)})  \ee{\ii \beta_d \alpha_{\sigma(d)}}\ket{0} \right\|^2
			= \left| \sum_{\sigma \in \Sigma_N} \prod_{d \in \mathcal{D}} \mathcal{U}_{d \sigma(d)} f(\beta_d-\beta_{\sigma(d)}) \ee{\ii \beta_d \alpha_{\sigma(d)}}\right|^2.
	\end{split}
\end{equation}
If the detector integration intervals $I(\{\beta_d\})$ are sufficiently small, Eq.~\eqref{eq:supp:probability_frequency} consequently reduces to Eq.~\eqref{eq:probability_ideal_measurement_general_state} given in the main paper.

\section{Inner-mode resolved correlation symmetries of interfering nonidentical photons}

\subsection{Parity invariance}

Let us assume that the photon spectra are symmetric around a common central value $\beta_0$.
In this case, we can write all expressions in terms of the differences $\nu_d \defeq \beta_d - \beta_0$ and Eq.~\eqref{eq:conjugate_symmetry_spectral_amplitude} follows immediately since
\begin{align}
	\mathcal{B}^{*}_{\sigma}(\{\nu_d\}) &= \prod_{d=1}^{3} f_{\sigma(d)}(\nu_d) \ee{-\ii (\beta_0+\nu_d) \alpha_{\sigma(d)}} = \ee{-\ii 2\beta_0(\alpha_{1}+\alpha_{2}+\alpha_{3})} \prod_{d=1}^{3} f_{\sigma(d)}(\nu_d) \ee{\ii(\beta_0- \nu_d) \alpha_{\sigma(d)}} \\
	&= \ee{-\ii 2\beta_0(\alpha_{1}+\alpha_{2}+\alpha_{3})} \prod_{d=1}^{3} f_{\sigma(d)}(-\nu_d) \ee{\ii(\beta_0- \nu_d) \alpha_{\sigma(d)}} \\
	&= \ee{-\ii 2\beta_0(\alpha_{1}+\alpha_{2}+\alpha_{3})} \mathcal{B}_{\sigma}(\{-\nu_d\}).
	\label{eq:supp:symmetric_spectra}
\end{align}
Combining this with the condition $\conj{\mathcal{A}}_{\sigma} = \abs{\mathcal{A}_{\sigma}} \ee{-\ii\varphi} = \ee{-2\ii\varphi} \mathcal{A}_{\sigma} \ \forall \sigma$ yields the parity symmetry
\begin{equation}
	\label{eq:supp:parity_invariance}
	\begin{split}
		P^{(\mathcal{D},\mathcal{S})}_{\{-\nu_{d}\},\{\alpha_s\}}
		&= \abs[\Big]{\sum_{\sigma} \mathcal{A}_{\sigma} \mathcal{B}_{\sigma}(\{-\nu_d\},\{\alpha_s\})}^2
		= \abs[\Big]{\sum_{\sigma} \mathcal{A}_{\sigma} \conj{\mathcal{B}}_{\sigma}(\{\nu_d\},\{\alpha_s\})}^2
		\\
		&= \abs[\Big]{\sum_{\sigma} \conj{\mathcal{A}}_{\sigma} \mathcal{B}_{\sigma}(\{\nu_d\},\{\alpha_s\})}^2
		= \abs[\Big]{\sum_{\sigma} \ee{-2\ii\varphi}\mathcal{A}_{\sigma} \mathcal{B}_{\sigma}(\{\nu_d\},\{\alpha_s\})}^2
		\\
		&= P^{(\mathcal{D},\mathcal{S})}_{\{\nu_d\},\{\alpha_s\}}
	\end{split}
\end{equation}
discussed in the main paper.

\subsection{Symmetries under permutations of the detected inner-mode values \texorpdfstring{$\beta_d$}{betad}}
We investigate how the probabilities for correlated detection in Eq.~\eqref{eq:probability_ideal_measurement_general_state} of the main paper behave under a permutation $\tau \in \Sigma_N$ of the arguments $\beta_d$.
From the definition of the spectral amplitudes $\mathcal{B}_{\sigma}$, we obtain
\begin{equation}
	\begin{split}
		\mathcal{B}_{\sigma}(\{\beta_{\tau(d)}\},\{\alpha_s\})
		&= \prod_{d=1}^{N} f_{\sigma(d)}(\beta_{\tau(d)}-\beta_{\sigma(d)})\ee{\ii \beta_{\tau(d)}\alpha_{\sigma(d)}}
		= \prod_{d=1}^{N} f_{\sigma(\tau^{-1}(d))}(\beta_d-\beta_{\sigma(\tau^{-1}(d))})\ee{\ii \beta_{d}\alpha_{\sigma(\tau^{-1}(d))}}
		\\
		&= \mathcal{B}_{\sigma\circ\tau^{-1}}(\{\beta_d\},\{\alpha_s\}).
	\end{split}
\end{equation}
Therefore, Eq.~\eqref{eq:permutation_of_detected_frequencies} in the main paper follows from Eq.~\eqref{eq:general_transformation_of_innermodes} in the main paper as
\begin{align}
	\label{eq:supp:probability_multiphoton_amplitudes_permuted}
	P^{(\mathcal{D},\mathcal{S})}_{\{\beta_{\tau(d)}\},\{\alpha_s\}} =
	\Big| \sum_{\sigma\in \Sigma_N}\mathcal{A}_{\sigma} \mathcal{B}_{\sigma\circ\tau^{-1}}(\{\beta_d\}) \Big|^2 = \Big| \sum_{\sigma\in \Sigma_N}\mathcal{A}_{\sigma\circ \tau} \mathcal{B}_{\sigma}(\{\beta_{d}\}) \Big|^2.
\end{align}
Here, we could relabel $\sigma\circ \tau^{-1}\rightarrow \sigma$ since, when summing over all permutations $\sigma\in \Sigma_N$, also $\sigma\circ\tau^{-1}$ covers all permutations in $\Sigma_N$.

\subsection{Mirror symmetry for tritters}
Here, we consider the case three photons ($N=3$), in which a mirror symmetry with respect to the plane (we again define $\nu_d=\beta_d-\beta_0$)
\begin{equation*}
	\left( \begin{array}{c}
			1 \\ 1 \\ 1
	\end{array} \right) \cdot
	\left( \begin{array}{c}
			\nu_1 \\ \nu_2 \\ \nu_3
\end{array} \right) = 0,
\end{equation*}
depicted as a light blue plane in Fig.~\ref{fig:correlation_landscape} in the main paper, appears for the highly symmetric state in which all photons share the \emph{same Gaussian distribution with the same central value $\beta_0$}.
This symmetry is \emph{independent of the interferometer transformation}.
Indeed, under the assumption that the inner-mode distributions of all single-photon pulses are identical Gaussian functions
\begin{equation}
	\label{eq:gaussian_frequency_spectra}
	f_s(\nu) = f(\nu) = \frac{\sqrt{\delta\beta}}{(2\pi \Delta\beta^2)^{1/4}}\exp\Big(-\frac{\nu^2}{4 \Delta\beta^2}\Big),
\end{equation}
the spectral amplitudes $\mathcal{B}_{\sigma}$ take the form
\begin{equation}
	\label{eq:supp:mirror_symmetry}
	\mathcal{B}_{\sigma}(\{\nu_d\},\{\alpha_s\}) = \Big( \frac{\sqrt{\delta\beta}}{(2\pi \Delta\beta^2)^{1/4}} \Big)^{3} \exp\Big( - \frac{\sum_{d=1}^3\nu_d^2}{4 \Delta\beta^2} \Big) \ee{\ii \sum_{d}\nu_d \alpha_{\sigma(d)}}.
\end{equation}
In order to investigate the effect of the mirror operation on the amplitudes $\mathcal{B}_{\sigma}$, an expression for the corresponding unitary transformation matrix acting on the vector $(\beta_1,\beta_2,\beta_3)^T$ is needed.
It can be found easily by noting that the operation can be divided into three steps:
First, a rotation is applied which maps the normal vector $\vec{n}=(1,1,1)^T/\sqrt{3}$ to the unit vector $\vec{e}_3=(0,0,1)^T$ in $\beta_3$ direction.
Then, the mirror operation can simply be described as the inversion of the sign of the inner-mode value $\beta_3$.
To complete the mirroring, the initial rotation then has to be inverted.

The rotation mapping $\vec{n}$ to $\vec{e}_3$ can be described as a rotation around the axis $\vec{u}_{\vec{n}}=(1,-1,0)/\sqrt{2}$ by the angle $\theta_{\vec{n}}=\arccos(\vec{n}\cdot\vec{e}_3)$.
With the help of the rotation matrix
\begin{equation}
	\label{eq:supp:general_rotation_matrix}
	R(\vec{u},\theta) =
	\begin{pmatrix}
		\cos\theta + u_1^{2}(1-\cos\theta) & u_1 u_2 (1-\cos\theta)-u_3 \sin\theta & u_1 u_3(1-\cos\theta) + u_2 \sin\theta \\
		u_2 u_1 (1-\cos\theta) + u_3 \sin\theta & \cos\theta + u_2^{2} (1-\cos\theta) & u_2 u_3 (1-\cos\theta) - u_1 \sin\theta \\
		u_3 u_1 (1-\cos\theta) - u_2 \sin\theta & u_3 u_2 (1-\cos\theta) + u_1 \sin\theta & \cos\theta + u_3^2 (1-\cos\theta)
	\end{pmatrix}.
\end{equation}
describing a general rotation by an angle $\theta$ around an axis $\vec{u}$, we can finally express the mirror operation as
\begin{equation}
	\label{eq:supp:mirror_transformation}
\begin{pmatrix}
\nu_1 \\ \nu_2 \\ \nu_3
\end{pmatrix} \rightarrow
\begin{pmatrix}
	\tilde{\nu}_1 \\ \tilde{\nu}_2 \\ \tilde{\nu}_3
\end{pmatrix} =
R(\vec{u}_{\vec{n}},-\theta_{\vec{n}})
\begin{pmatrix*}[r]
	1 & 0 & 0 \\
	0 & 1 & 0 \\
	0 & 0 & -1
\end{pmatrix*}
R(\vec{u}_{\vec{n}},\theta_{\vec{n}})
\begin{pmatrix}
	\nu_1 \\ \nu_2 \\ \nu_3
\end{pmatrix}
=
\frac{1}{3}\begin{pmatrix*}[r]
	1 & -2 & -2 \\
	-2 & 1 & -2 \\
	-2 & -2 & 1
\end{pmatrix*}
\begin{pmatrix}
	\nu_1 \\ \nu_2 \\ \nu_3
\end{pmatrix}.
\end{equation}
It is obvious that the argument of the first exponential in Eq.~\eqref{eq:supp:mirror_symmetry} is invariant (due to its spherical symmetry) under this transformation while the second exponential becomes
\begin{equation*}
	\exp\Big(\ii{\sum_{d=1}^3 \tilde{\nu}_d \alpha_{\sigma(d)}}\Big) = \exp\Big({-\ii\frac{2}{3} \sum_{d=1}^3 \nu_d \sum_{s=1}^3 \alpha_{s}}\Big) \exp\Big(\ii{\sum_{d=1}^3 \nu_d \alpha_{\sigma(d)}}\Big),
\end{equation*}
leading to
\begin{equation*}
	\mathcal{B}_{\sigma}(\{\tilde{\nu}_d\},\{\alpha_s\}) = \exp\Big({-\ii\frac{2}{3} \sum_{d=1}^3 \nu_d \sum_{s=1}^3 \alpha_{s}}\Big) \mathcal{B}_{\sigma}(\{\nu_d\},\{\alpha_s\}).
\end{equation*}
Therefore, the mirror transformation only introduces a complex phase factor which is equal for all multiphoton amplitudes and consequently does not contribute to the correlations, Eq.~\eqref{eq:probability_ideal_measurement_general_state} in the main text, i.e.
\begin{equation}
	P^{(\mathcal{D},\mathcal{S})}_{\{\tilde{\nu}_d\},\{\alpha_s\}} = P^{(\mathcal{D},\mathcal{S})}_{\{\nu_d\},\{\alpha_s\}}.
\end{equation}

\section{Multiphoton entanglement features}
\label{sub:entanglement_of_post_selected_states}

\subsection{\texorpdfstring{Emergence of the whole class of $W$-states}{Emergence of the whole class of W-states}}

Choosing a computational basis $H$, $V$ for the polarization, a single-photon pulse at the input port $s$ with frequency distribution $\xi_s(\omega)$, initial time $t_{s}$, and polarization $\lambda_s \in \{H,V\}$ is described by the state
\begin{equation}
	\label{eq:supp:single_photon_state_polarization}
	\ket{1;\xi_s,t_{s},\lambda_s}_s \defeq \int_{0}^{\infty} \d{\beta} f_s(\beta) \ee{\ii \beta \alpha_{s}} \op{a}^{\dagger}_{s,\lambda_s}(\beta)\ket{0}_s,
\end{equation}
where $f_s = \xi_s,\mathcal{F}[\xi_s]$ if $\beta=\omega,t$.
Using the polarized single-photon input pulses $\ket{1;\xi_s,\omega_s,t_{s},\lambda_s}_s$, defined in Eq.~\eqref{eq:singlephotonstate_polarization}, the input state of an $N$-port interferometer is
\begin{equation}
	\label{eq:supp:input_polarization}
	\ket{\psi_{\text{in}}} \defeq \bigotimes_{s=1}^{N}\ket{1;\xi_s,\omega_s,t_{s},\lambda_s}_s.
\end{equation}
By rewriting the input-port creation operators $\op{a}_s$ in terms of the output-port creation operators $\op{a}_d$ as
\begin{equation}
	\label{eq:supp:linear_transformation}
	\op{a}^{\dagger}_{s,\lambda_s}(\beta) = \sum_{d} \mathcal{U}_{ds} \op{a}^{\dagger}_{d,\lambda_s}(\beta),
\end{equation}
the propagation of the input state $\ket{\mathcal{S}}$, defined in Eq.~\eqref{eq:StateDefinition} of the main paper, to the output of the interferometer leads to
\begin{equation}
	\label{eq:supp:output_state_polarization}
	\ket{\psi_{\text{out}}} = \prod_{s=1}^{N}\sum_{d_s=1}^N \mathcal{U}_{d_ss} \int \d{\beta_s} f_s(\beta_{s})\ee{\ii\beta_s \alpha_{s}}\op{a}^{\dagger}_{d_s,\lambda_s}(\beta_s) \ket{0}.
\end{equation}
In the following, we will only consider events, in which a photon is detected at each of the output ports $d=1,\dots,N$.
The corresponding part of the output state reads
\begin{align}
	\ket{\psi^{(1\dots N)}_{\text{out}}} &= \sum_{\sigma\in\Sigma_N}\prod_{d=1}^{N} \mathcal{U}_{d\sigma(d)} \int \d{\beta_d} f_{\sigma(d)}(\beta_{d})\ee{\ii\beta_{d} \alpha_{\sigma(d)}}\op{a}^{\dagger}_{d,\lambda_{\sigma(d)}}(\beta_d) \ket{0} \\
	\label{eq:supp:output_state_polarization_no_bunching}
	&= \sum_{\sigma\in\Sigma_N} \int \prod_{d=1}^{N}\d{\beta_d} \mathcal{A}_{\sigma}\prod_{d=1}^{N}f_{\sigma(d)}(\beta_{d})\ee{\ii\beta_{d} \alpha_{\sigma(d)}} \ket{\lambda_{\sigma(1)},\dots,\lambda_{\sigma(N)};\{\beta_d\}}.
\end{align}

\subsection{Three-photon W-state}
Let us now assume that $N=3$ and that all three photons have equal spectral distributions, $\xi_{s}(\omega)=\xi(\omega) \ \forall s=1,2,3$, and that $\lambda_1=\lambda_2=H$ and $\lambda_3=V$. We can then rewrite Eq.~\eqref{eq:supp:output_state_polarization_no_bunching} in the form
\begin{align}
	\ket{\psi^{(123)}_{\text{out}}}
	&= \int \d{\beta_1}\int \d{\beta_2} \int \d{\beta_3} \prod_{d=1}^{3}f(\beta_{d})\sum_{\sigma\in\Sigma_3} \mathcal{A}_{\sigma} \ee{\ii\beta_{d} \alpha_{\sigma(d)}} \ket{\lambda_{\sigma(1)},\lambda_{\sigma(2)},\lambda_{\sigma(3)};\{\beta_d\}} \\
	&\begin{multlined}[b]
	= \int \d{\beta_1}\int \d{\beta_2} \int \d{\beta_3} \prod_{d=1}^{3}f(\beta_{d})
		\Bigg[ \smashoperator[r]{\sum_{\sigma\in\{(13)(2),(132)\}}} \mathcal{A}_{\sigma} \ee{\ii\beta_{d} \alpha_{\sigma(d)}} \ket{V,H,H;\{\beta_d\}} \\
		+\smashoperator[r]{\sum_{\sigma\in\{(23)(1),(123)\}}} \mathcal{A}_{\sigma} \ee{\ii\beta_{d} \alpha_{\sigma(d)}} \ket{H,V,H;\{\beta_d\}}
		+ \smashoperator[r]{\sum_{\sigma\in\{(12)(3),(1)(2)(3)\}}} \mathcal{A}_{\sigma} \ee{\ii\beta_{d} \alpha_{\sigma(d)}} \ket{H,H,V;\{\beta_d\}}  \Bigg]
	\end{multlined} \\
	\label{eq:supp:output_state_no_bunching_symmetric_tritter}
	& \defeqinv \int \d{\beta_1}\int \d{\beta_2} \int \d{\beta_3} \prod_{d=1}^{3}f(\beta_{d}) \ee{\ii \beta_3 \alpha_d} \Big( \tilde{a}_{\{\beta_d\}} \ket{V,H,H;\{\beta_d\}} + \tilde{b}_{\{\beta_d\}} \ket{H,V,H;\{\beta_d\}} + \tilde{c}_{\{\beta_d\}} \ket{H,H,V;\{\beta_d\}} \Big),
\end{align}
Inserting the interferometer multiphoton amplitudes $\mathcal{A}_{\sigma}$ for the symmetric tritter $\mathcal{U}_{ds} = \exp(\ii2\pi d s/3)/\sqrt{3}$ the coefficient $\tilde{a}$ evaluates as
\begin{align}
	\tilde{a}_{\{\beta_d\},\{\alpha_s\}} &= \ee{-\ii \beta_3(\alpha_1+\alpha_2+\alpha_3)} \Big[ 3^{-3/2} \ee{-\ii 2\pi/3} \ee{\ii (\beta_1  \alpha_3 + \beta_2 \alpha_1 + \beta_3 \alpha_2)}  + 3^{-3/2} \ee{+\ii 2\pi/3} \ee{\ii (\beta_1 \alpha_3 + \beta_2 \alpha_2 +\beta_3 \alpha_1)} \Big] \\
	&=  3^{-3/2} \Big[ \ee{-\ii 2\pi/3} \ee{\ii \big( (\beta_1-\beta_3) \alpha_3 + (\beta_2-\beta_3) \alpha_1 \big)}  + \ee{+\ii 2\pi/3} \ee{\ii \big( (\beta_1-\beta_3) \alpha_3 + (\beta_2-\beta_3) \alpha_2 \big)} \Big],
\end{align}
or
\begin{align}
	\label{eq:supp:definition_wstate_like_coefficients_a}
	\tilde{a}=\tilde{a}_{\{\beta_d\},\{\alpha_s\}}
	&= \frac{2}{3^{3/2}} \ee{\ii(\beta_1-\beta_3) \alpha_3} \ee{\ii \frac{1}{2}(\beta_2-\beta_3)(\alpha_1+\alpha_2)} \cos\Big( \frac{1}{2}(\beta_2-\beta_3)(\alpha_1-\alpha_2)- \frac{2\pi}{3} \Big).
	\intertext{Analogously, the remaining coefficients are}
	\label{eq:supp:definition_wstate_like_coefficients_b}
	\tilde{b} =\tilde{b}_{\{\beta_d\},\{\alpha_s\}}
	&= \frac{2}{3^{3/2}} \ee{\ii \frac{1}{2}(\beta_1-\beta_3)(\alpha_{1}+\alpha_{2})} \ee{\ii (\beta_2-\beta_3) \alpha_{3}} \cos \left( \frac{1}{2}(\beta_1-\beta_3)(\alpha_{1}-\alpha_{2}) + \frac{2\pi}{3}\right) \\
	\label{eq:supp:definition_wstate_like_coefficients_c}
	\tilde{c}=\tilde{c}_{\{\beta_d\},\{\alpha_s\}}
	&= \frac{2}{3^{3/2}}\ee{\ii \frac{1}{2}((\beta_1-\beta_3)+(\beta_2-\beta_3))(\alpha_{1}+\alpha_{2})} \cos \left( \frac{1}{2}((\beta_1-\beta_3)-(\beta_2-\beta_3))(\alpha_{1}-\alpha_{2}) - \frac{2\pi}{3}\right).
\end{align}
Since the coefficients only depend on the differences in the inner-mode values $\beta_d$ we made the arbitrary choice to write them as functions of the pair of differences $\beta_1-\beta_3$ and $\beta_2-\beta_3$. Equivalently, we could have chosen $\beta_1-\beta_2$ and $\beta_3-\beta_2$ or $\beta_2-\beta_1$ and $\beta_3-\beta_1$.
Consequently, for any given set of detected inner-mode values $\{\beta_d\}$ the measured state is
\begin{equation}
	\label{eq:supp:definition_wstate}
	\ket{W;\{\beta_d\}} \defeq a \ket{V,H,H;\{\beta_d\}} + b \ket{H,V,H;\{\beta_d\}} + c \ket{H,H,V;\{\beta_d\}}
\end{equation}
with the coefficients
\begin{align}
	\label{eq:supp:aomega}
	a = a_{\{\beta_d\},\{\alpha_s\}} &= \frac{\tilde{a}}{|\tilde{a}|^2+|\tilde{b}|^2+|\tilde{c}|^2}, \\
	b = b_{\{\beta_d\},\{\alpha_s\}} &= \frac{\tilde{b}}{|\tilde{a}|^2+|\tilde{b}|^2+|\tilde{c}|^2}, \\
	\label{eq:supp:comega}
	c = c_{\{\beta_d\},\{\alpha_s\}} &= \frac{\tilde{c}}{|\tilde{a}|^2+|\tilde{b}|^2+|\tilde{c}|^2},
\end{align}
whose squared moduli are plotted for the exemplary case $\beta_1 - \beta_3 = -(\beta_2-\beta_3)$ in Fig.~\ref{fig:entanglement}a) in the main paper.

\subsection{\texorpdfstring{Inner-mode dependent entanglement in $W$-states}{Inner-mode dependent entanglement in W-states}}

The three-photon $W$-states are defined as \cite{Dur2000_Threequbitscan_S}
\begin{equation}
	\label{eq:supp:wstate_class}
	\ket{W} = a \ket{1}_A\ket{0}_B\ket{0}_C + b \ket{0}_A\ket{1}_B\ket{0}_C + c \ket{0}_A\ket{0}_B\ket{1}_C,
\end{equation}
	with $|a|^2+|b|^2+|c|^2=1$\footnote{Note that the definition of the $W$-state class in \cite{Dur2000_Threequbitscan_S} includes a fourth state component $\protect\ket{0}_A\protect\ket{0}_B\protect\ket{0}_C$. However, the entanglement between the three qubits is not influenced by this fourth component, as shown in  \cite{Dur2000_Threequbitscan_S}}.
	This class of states shows true tripartite-entanglement, i.e. is neither 2- nor 3-separable, as long as $a,b,c \neq 0$. In this class, there is no entanglement that is shared between all three particles at the same time, however each pair of particles is entangled. We can therefore quantify the entanglement of the states in Eq.~\eqref{eq:supp:wstate_class} using the concurrences between all three pairs of the subsystems $A$, $B$, and $C$. For example, the entanglement between subsystems $A$ and $B$ of a tripartite state $\rho_{ABC}$ can be quantified \cite{Wootters1998_Entanglementformationarbitrary,Coffman2000_Distributedentanglement_S} by using the reduced density matrix
	\begin{equation}
		\label{eq:supp:reduced_density_matrix}
		\rho_{AB} \defeq \tr_{C} \rho_{ABC},
	\end{equation}
	where $\tr_C$ denotes the partial trace over subsystem $C$.
	By using the Pauli spin matrix $\sigma_y$, we can define the matrix
	\begin{equation}
		\label{eq:supp:definition_auxiliary_matrix_concurrence}
		R_{AB} \defeq \sqrt{\sqrt{\rho_{AB}} (\sigma_y \otimes \sigma_y) \rho_{AB}^* (\sigma_y \otimes \sigma_y) \sqrt{\rho_{AB}} }
	\end{equation}
	whose eigenvalues $\lambda_1 \geq \lambda_2 \geq \lambda_3 \geq \lambda_4$ determine the concurrence
	\begin{equation}
		\label{eq:supp:concurrence_squared_definition}
		C_{AB} \defeq \max(0, \lambda_1-\lambda_2-\lambda_3-\lambda_4).
	\end{equation}
	From these definitions, we find for the class of states in Eq.~\eqref{eq:supp:wstate_class} that (see \cite{Dur2000_Threequbitscan_S})
	\begin{align}
		\label{eq:supp:concurrences_wstate_class}
		C_{AB} = 2 |a||b|; \quad C_{BC} = 2 |b||c|; \quad C_{AC} = 2|a||c|.
	\end{align}
	To quantify the tripartite entanglement, we use the average squared concurrence (the inequalities are proved in \cite{Dur2000_Threequbitscan_S})
	\begin{equation}
		\label{eq:supp:average_concurrence}
		E_{\text{av}} \defeq \frac{1}{3}(C_{AB}^2 + C_{BC}^2 + C_{AC}^2) \leq \frac{4}{9} \quad\text{   (equality for $\abs{a}=\abs{b}=\abs{c}$)}
	\end{equation}
	and the minimal squared concurrence
	\begin{equation}
		\label{eq:supp:minimal_concurrence}
		E_{\text{min}} \defeq \min(C_{AB}^2,C_{BC}^2,C_{AC}^2) \leq \frac{4}{9} \quad \text{   (equality for $\abs{a}=\abs{b}=\abs{c}$)}.
	\end{equation}

	By explicitly writing the concurrences in Eq.~\eqref{eq:supp:concurrences_wstate_class} in terms
	of the coefficients in Eqs.~\eqref{eq:supp:aomega}-\eqref{eq:supp:comega} and substituting their expressions into
	Eqs.~\eqref{eq:supp:average_concurrence} and \eqref{eq:supp:minimal_concurrence} we find the average concurrence and the minimum concurrence, respectively, as a function of the detected inner-mode values at the output of a symmetric tritter, as reported in the main paper.

\end{document}